\documentstyle[12pt]{article}
\begin{document}

\rightline{DFPD 95/TH/11}
\rightline{DFSA 95/TH/5}
\rightline{March 1995}

\vspace{0.2cm}

{\begin{center}
{\large \bf DYNAMICS OF SQUEEZING FROM \\

\vspace{0.2cm}

            GENERALIZED COHERENT STATES} \\

\vspace{1.3cm}

Salvatore De Martino \footnote{Electronic Mail:
demartino@vaxsa.dia.unisa.it}$^{\ast}$,
Silvio De Siena \footnote{Electronic
Mail: desiena@vaxsa.dia.unisa.it}$^{\ast}$,
and Fabrizio Illuminati \footnote{Electronic Mail:
illuminati@mvxpd5.pd.infn.it}$^{\S}$

\vspace{0.4cm}

$^{\ast}$ {\it Dipartimento di Fisica, Universit\`{a} di Salerno, \\
     and INFN, Sezione di Napoli, Gruppo collegato di Salerno, \\
     84081 Baronissi, Italia}

\vspace{0.2cm}

$^{\S}$ {\it Dipartimento di Fisica ``G. Galilei",
Universit\`{a} di Padova, \\
and INFN, Sezione di Padova, 35131 Padova, Italia}

\end{center}}

\vspace{0.5cm}

{\begin{center} \large \bf Abstract \end{center}}
We extend the definition of generalized coherent states
to include the case of time-dependent dispersion.
We introduce a suitable operator providing displacement
and dynamical rescaling from an arbitrary ground state.
As a consequence, squeezing is naturally embedded in
this framework, and its dynamics is
ruled by the evolution equation for the dispersion.
Our construction provides a displacement-operator method
to obtain the squeezed states of arbitrary systems.

\vspace{0.4cm}


\newpage

{\it Introduction}. Coherent states are the quantum states
that are closest to a classical, localized time-evolution; after
the pioneering work by Schr\"odinger \cite{schroedinger}, they
were discovered and their structure thoroughly clarified
in the modern language of quantum field theory by Glauber,
Klauder, and Sudarshan \cite{glauber}, \cite{klaudersuda}.

Besides their conceptual
relevance to the understanding of basic features of quantum
mechanics, they are by now an indispensable mathematical
tool in many fields of theoretical physics, ranging from
quantum field theory to statistical mechanics
\cite{klauderskagilmore}, \cite{perelomov}.

Their extension, squeezed states \cite{yuen},
have come to play an increasingly important role
in the last decade; beyond the traditional context of
quantum electrodynamics and quantum optics \cite{kim},
they also appear to be of special interest in the theory of
quantum nondemolition measurements applied to gravitational
wave detection \cite{braginskii}.

In this note we address the problem of constructing
and deriving the dynamical properties of squeezed states
for arbitrary systems.

The results that we are going to present stem from a new
approach to  coherent states that is
based on Nelson stochastic mechanics, a
quantization scheme \cite{nelson67},
whose motivation arises from a cross-breeding of ideas
and methods of euclidean quantum field theory
\cite{guerraruggiero}, the theory of stochastic differential
equations \cite{blanchardparisi}, and the theory of
stochastic optimal control \cite{guerramorato}.

New results and insights can then be obtained
by looking at quantum coherence in terms of
general properties of classical diffusion processes.
This approach has been carried out gradually.

We first derived the standard
harmonic-oscillator coherent and squeezed states
as the Nelson diffusions minimizing the osmotic uncertainty
relations of stochastic mechanics \cite{demartino94}.

Next, we introduced a class of generalized coherent states
as the Nelson diffusions with classical current velocity and
wave-like propagating osmotic velocity \cite{demartinobis94}.
These coherent states follow a classical motion in generic
time-dependent potentials without spreading of the wave packet.
 The evolution
is controlled by a feedback mechanism that allows the packets
to remain coherent through a continuous dynamical readjustment.

The connection with standard operatorial languages was then
provided by showing that the new class of states is generated
letting the Glauber displacement operator $\hat{D}(\alpha)$ act
on the ground states of arbitrary potentials in the coordinate
representation \cite{demartinoter94}. In this way we derived
a complete dynamical description of the displacement-operator
generalized coherent states.

In this letter we extend the scheme previously developed to include
the case of time-dependent dispersion $\Delta q$. By resorting to
the stochastic framework we are able to derive the desired
evolution equation for $\Delta q$, as well as the general form
of the wave functions associated to this new class of
coherent states.

A suitable displacement operator is then introduced
in the coordinate representation in order to obtain
these states in standard operatorial language: they are
displacement-operator generalized coherent states with
time-dependent dispersion.

In fact, the displacement operator acts as the product
of two distinct mappings: the ordinary Glauber displacement
operator $\hat{D}(\alpha)$ and a dynamical rescaling
operator, namely a dynamical ``squeeze" operator.

Squeezing is then naturally embedded in this scheme,
and the evolution equation for $\Delta q$ yields also the
dynamical equation controlling the time-evolution of
squeezing.

At the same time, the above construction provides
a natural extension of the displacement-operator
method to define a class of squeezed states for
arbitrary potentials.

{\it Stochastic mechanics}. We shall quickly
review the basic ingredients of the stochastic formulation
of quantum mechanics that will be needed in the following.

This quantization procedure rests on
two basic prescriptions; the first one, kinematical, promotes
the configuration of a classical system to a conservative
diffusion process with diffusion coefficient equal
to $\hbar/2m$.

If we denote by $q(t)$ the configurational variable for a point
particle with mass $m$, this prescription reads
\begin{equation}
dq(t) = v_{(+)}(q(t),t)dt + \sqrt{\frac{\hbar}{2m}}
dw(t) \, , \; \; \; \; dt > 0 \, \, .
\end{equation}

\noindent In the above stochastic differential equation
$v_{(+)}$ is a (forward) drift field that is determined
by assigning the dynamics, and $w$ is the standard
Wiener process.

An intuitive manner to look at Eq. (1) is to consider it as
the appropriate quantum form of the classical kinematical
prescription: the Wiener process models
quantum fluctuations, just as in the theory of beams
dynamics in particle accelerators.

Under very general mathematical conditions, the diffusion
$q(t)$ admits the backward representation
\begin{equation}
dq(t)=v_{(-)}(q(t),t)dt + \sqrt{\frac{\hbar}{2m}}
dw^{\ast}(t)\, ,\; \;\;\; dt>0 \, \, ,
\end{equation}

\noindent where $w^{\ast}$ is a time-reversed Wiener process
and the backward drift is defined by the relation
\begin{equation}
 v_{(-)} \; = \; v_{(+)} \,
- \, \frac{\hbar}{2m}\nabla\ln\rho \, .
\end{equation}

In Eq.(3) above $\rho(x,t)$ denotes the probability density
associated to the process $q(t)$. It is useful to introduce
the hydrodynamic representation in terms of the osmotic
and current velocity, defined respectively by
\begin{equation}
u(x,t)= \frac{v_{(+)}-v_{(-)}}{2} \, ,
\end{equation}

and
\begin{equation}
v(x,t)=\frac{v_{(+)} + v_{(-)}}{2} \, .
\end{equation}

In the hydrodyanmic picture, the Fokker-Planck
equation associated to Eqs. (1)-(2) takes the form of
a continuity equation:
\begin{equation}
\partial_{t}\rho \; = \; -\nabla(\rho v) \, .
\end{equation}

The dynamical prescription is introduced by defining the
mean regularized classical action $A$. In the hydrodynamic
Eulerian picture it is a functional of the couple $(\rho,v)$:
\begin{equation}
A \; = \; \int_{t_{a}}^{t_{b}} \left[
\frac{m}{2}(v^2 - u^2) -\Phi \right] \rho d^{3}x \, ,
\end{equation}

\noindent where $\Phi(x,t)$ denotes the external potential.

The equations of motion are then obtained by extremizing $A$
against smooth variations $\delta \rho$, $\delta v$ vanishing
at the boundaries of integration, with the continuity equation
taken as a constraint.

After standard calculations one obtains
\begin{equation}
\partial_{t}v + (v\cdot \nabla )v - \frac{\hbar^2}{4m^2}
\nabla\left( \frac{\nabla^2\sqrt{\rho}}{\sqrt{\rho}} \right) =
 -\nabla\Phi \, ,
\end{equation}

\noindent with the current velocity fixed to be a gradient field
at all points where $\rho > 0$: $v = \nabla S/m$,
where $S(x,t)$ is a scalar function.

Defining the wave function $\Psi (x,t)$ for
a generic quantum state in the hydrodynamic form
$\Psi = \sqrt{\rho}\exp \left[ iS/\hbar \right] $, one
immediately has that the continuity equation together
with the dynamical equation Eq. (8) are equivalent to
the Schr\"{o}dinger equation.

The space integral of Eq. (8) yields the Hamilton-Jacobi-Madelung
equation. It is useful for
what follows to write this equation in the form
\begin{equation}
\partial_{t}S + \frac{m}{2}v^{2} - \frac{m}{2}u^{2} - \frac{\hbar}{2}
\partial_{x}u \; = \; - \Phi \; .
\end{equation}

The correspondence between expectations and correlations defined
in the stochastic and in the canonic pictures are
\[
\langle \hat{q} \rangle = E(q) \, , \; \; \; \; \; \; \; \; \; \; \;
\; \langle \hat{p} \rangle = mE(v) \, ,
\]
\begin{equation}
\end{equation}

\[
\Delta \hat{q} = \Delta q \, , \; \; \; \; \; \;
(\Delta \hat{p})^{2} = m^{2}[(\Delta u)^{2} + (\Delta v)^{2}] \, ,
\]

\noindent The following chain inequality holds:
\begin{equation}
(\Delta \hat{q})^{2} (\Delta \hat{p})^{2} \, \geq \,
m^{2}(\Delta q)^{2} (\Delta u)^{2} \, \geq \,
\frac{{\hbar}^{2}}{4} \, .
\end{equation}

In the above relations $\hat{q}$ and $\hat{p}$ denote
the position and momentum observables in the Schr\"odinger
picture, $\langle \cdot \rangle$ denotes the expectation value
of the operators in the given state $\Psi$, $E(\cdot)$ is the
expectation value of the
stochastic variables associated in the Nelson picture to the
state $\{\rho, v\}$, and $\Delta(\cdot)$ denotes the root mean
square deviation.

The inequalities Eq. (11), i.e. the osmotic
uncertainty relation and its equivalence with the
momentum-position uncertainty were proven in
Ref.\cite{demartino82}.

{\it Harmonic-oscillator coherent and squeezed states}.
Saturation of the osmotic uncertainty relation Eq. (11) leads to
the definition of Glauber coherent states in the stochastic
picture \cite{demartino94}: they are Nelson diffusions of
constant dispersion $\Delta q$, and with classical current
velocity and linear osmotic velocity:
\begin{equation}
v \, = \, \langle v \rangle \; \; \; \; \; \; \; \; \; \;
u \, = \, - \frac{\hbar}{2m\Delta q} \xi \, .
\end{equation}

We have now denoted the stochastic expectations with the same
symbol used for quantum ones, and we have introduced the
adimensional variable $\xi = (x - \langle q \rangle )/\Delta q$.

The harmonic-oscillator squeezed states are instead Nelson
diffusions with time-varying $\Delta q$, linear $u$ of the
form Eq. (12) and current velocity of the form \cite{demartino94}:
\begin{equation}
v \, = \, \langle v \rangle \, + \, \xi \frac{d}{dt}\Delta q \, .
\end{equation}

The last term in Eq. (13) is responsible for the quantum
anticommutator term appearing in the phase of the squeezed
wave packets.

Of course, both the coherent and the squeezed states Eqs. (12)-(13)
follow the classical motion
\begin{equation}
\frac{d}{dt}(m \langle v \rangle ) \, = \, - \nabla \Phi (x,t)
|_{x=\langle q \rangle } \, ,
\end{equation}

\noindent where,  $\langle v \rangle =
d\langle q \rangle /dt$, a well known classical property of quantum
and stochastic expectations.

Harmonic-oscillator coherent states can also be obtained in
a stochastic variational approach by extremizing the osmotic
uncertainty product against smooth variations of the density
$\rho$ and of the current velocity $v$ \cite{illuminati95}.
The possibility of extending
this approach to study local minimum
uncertainty behaviors in non harmonic
systems is currently being investigated, see Ref.
\cite{illuminati95}.

{\it Generalized coherent states}.
{}From a dynamical point of view a coherent state is a wave packet
whose centre follows a classical motion and whose dispersion
is either constant or controlled in its time-evolution (squeezing).

In quantum mechanics the dynamics of mean values obeys Ehrenfest
theorem: as a consequence, the coherent evolution Eq. (14) is strictly
satisfied if
\begin{equation}
\langle \nabla \Phi (x,t) \rangle \;
= \; \nabla \Phi (x,t)|_{x= \langle q \rangle} \, .
\end{equation}

In the case of quadratic potentials the above constraint
is authomatically satisfied for any quantum state.

For other generic potentials $V(x,t)$ Eq. (15) in general
cannot be satisfied. However, in stochastic mechanics a
particular choice of the current velocity selects an entire
class of osmotic velocities, i.e. of quantum states.

In particular, we showed that, in the case of constant $\Delta q$,
the choice $v = \langle v \rangle$ does not fix $u$ to be only
of the standard Glauber form Eq. (12), but yields instead a whole
class of quantum states with osmotic velocities of
the wave-like propagating form $u=G(\xi)/\Delta q$,
with $G$ arbitray function \cite{demartinobis94}.

What are the properties of this class of states? They are no
more Heisenberg minimum uncertainty states (the
latter are recovered choosing $G=-\hbar \xi/2m$).
However, they obey the constraint Eq. (15) apart, at most,
a constant.

Moreover, they can be obtained in the coordinate representation
by applying the displacement operator on the ground state wave
functions of arbitrary configurational potentials
$V_{0}(x)$ \cite{demartinoter94}.

They are then generalized coherent states of the displacement
operator, following classical motion with constant dispersion;
we showed that this is possible because they obey Schr\"{o}dinger
equation in time-dependent potentials $V(x,t)$ with
a dynamical feedback mechanism allowing the wave
packet to remain coherent.

These states exhaust the class of possible ones
obeying a generalized Glauber condition Eq. (15).
In fact, the harmonic oscillator is the trivial
instance  for which
the feedback mechanism disappears \cite{demartinoter94}.

In conclusion, we first selected via stochastic mechnanics
the quantum systems, beyond the harmonic oscillator,
that can obey constraint Eq. (15) and we then showed that
these systems are associated with the displacement-operator
coherent states, and derived their dynamical properties.

{\it Time-dependent dispersion: generalized ``dynamical"
coherent states}.
We now proceed to build  the case of
time-dependent dispersion, that is we consider the more general
form Eq. (13) for the current velocity of minimum uncertainty.

The generalized coherent states that we expect to select by taking
the choice Eq. (13) for the current velocity would obviously be
states following two coupled dynamical equations, Eq. (14) for
the wave packet centre, and an evolution equation for the dispersion
$\Delta q$, as in the Harmonic case.

We proceed as follows. By inserting Eq. (13) in the continuity
equation we are left with the  equation
\begin{equation}
\partial_{t} \rho \; = \; v \nabla \rho - \frac{1}{\Delta q}
\frac{d}{dt}\Delta q \, ,
\end{equation}

\noindent whose general solution is function only of $\xi =
(x- \langle q \rangle )/\Delta q$, and reminding that
$\rho$ must be non negative it can be cast in the form
\begin{equation}
\rho \; = \; \exp \left[ \frac{R(\xi)}{\Delta q} \right] \, ,
\end{equation}

\noindent with $R$ any arbitrary function yielding
a normalizable probability density.
By Eqs. (3)-(4) one then has
\begin{equation}
u \; = \; \frac{1}{\Delta q} G(\xi) \, .
\end{equation}

As a consequence, we have again that a class of osmotic velocities
of wave propagating form is selected, with the arbitrary function
$G$ restricted only by the normalization requirement for $\rho$.

Inserting now the current velocity Eq. (13) in the equation of
motion Eq. (8), by Eq. (18) one has
\begin{equation}
-m\xi \frac{d\Delta q}{dt}
+\frac{m}{2} \nabla u^2 + \frac{\hbar}{2} \nabla^2 u \;
= \; \nabla \Phi
-\langle \nabla \Phi \rangle \, ,
\end{equation}

\noindent where we exploited Ehrenfest theorem $d\langle v \rangle
/dt = -\langle \nabla \Phi \rangle /m$.

Letting $x=\langle q\rangle$ (i.e. $\xi =0$)
in Eq. (19), one has
\begin{equation}
\nabla {\Phi}\mid_{x=\langle q \rangle } \, - \,
\langle \nabla \Phi \rangle \; = \;
\frac{m}{2}\nabla u^2\mid_{\xi =0} + \frac{\hbar}{2}\nabla^2u\mid_{
\xi =0} \; .
\end{equation}

This is the same relation previously established in the case $v=
\langle v \rangle $ (time-independent $\Delta q$);
the right hand side is obviously either constant
or zero except for singular potentials: in these cases
$u$ diverges in $\xi = 0$. Explicit examples and applications
to non singular potentials, as well as a careful analysis
of the singular cases will be discussed in detail
elsewhere \cite{preparation}.

Next, we want to derive the phase $S$ and
the evolution equation for the
dispersion $\Delta q$. This is achieved by
exploiting Hamilton-Jacobi-Madelung equation, Eq. (9):
reminding that $v$ is the gradient field of $S$, Eq. (13)
implies
\begin{equation}
S \; = \; m\langle v \rangle x \, + \,
\frac{m}{2}\frac{(x - \langle q \rangle)^2}
{\Delta q}\frac{d\Delta q}{dt} \, + \, S_{0}(t) \; ,
\end{equation}

\noindent where $S_{0}(t)$ denotes the classical phase.
Inserting Eqs. (21), (13), and (18) in Eq. (9),
and taking the expectation value, we obtain the evolution
equation for $\Delta q$
\begin{equation}
\frac{m}{2}\Delta q \frac{d^2\Delta q}{dt^2}-\frac{m}{2}
{\langle v \rangle }^{2} + \frac{m}{2}\langle u^2 \rangle
\; = \; - \langle \Phi \rangle \, .
\end{equation}

\noindent By Eq. (18) for $u$ it is immediately seen that
\begin{equation}
\langle u^2 \rangle = \frac{K}{(\Delta q)^2} \; ,
\end{equation}

\noindent where $K=\int_{-\infty}^{\infty} G^{2}(\xi)d\xi$;
Eq. (22) is then the desired equation for the time-evolution
of the dispersion and is naturally coupled through the term
in $\langle v \rangle$ with the classical equation of motion
for the wave packet center $\langle q\rangle$, Eq. (14).

The general form of the wave function for such
states is readily obtained putting
together Eq. (17) for $\rho$ and Eq. (21) for $S$:
\begin{equation}
\Psi (x,t) \; = \; \exp \left[ \frac{R(\xi)}{\Delta q} \, + \,
\frac{i}{\hbar}\left( m\langle v \rangle x +
\frac{m}{2}\frac{(x - \langle q \rangle)^2}
{\Delta q}\frac{d\Delta q}{dt} + S_{0}(t) \right) \right] \, .
\end{equation}

We can rewrite this expression in more familiar terms by observing
that $m\Delta q d(\Delta q)/dt=\langle \{ \hat{Q},
\hat{P} \} \rangle /2$,
where $\hat{Q}=\hat{q} - \langle \hat{q} \rangle $, $\hat{P}=
\hat{p} - \langle \hat{p} \rangle $, and $\{ , \} $ denotes
the anticommutator. We thus have
\begin{equation}
\Psi (x,t) \; = \; \exp \left[ \frac{R(\xi)}{\Delta q} \, + \,
\frac{i}{\hbar}\left( \langle \hat{p} \rangle x +
\frac{\langle \{ \hat{Q}, \hat{P} \} \rangle }
{\left( 2\Delta q\right) ^{2}}(x - \langle \hat{q} \rangle)^2
+ S_{0}(t) \right) \right] \, .
\end{equation}

The above are nonstationary states with classical motion and
controlled time-dependent spreading, which evolve in the
time-dependent  potentials $V(x,t)$.

In the next section these generalized ``dynamical" coherent
states will be derived introducing Glauber displacement
operator and a proper ``squeeze" operator.

{\it Displacement operator: generalized squeezed states}.
It is well known that generalized coherent states can be obtained
extending the three different existing approaches to the definition
of the harmonic-oscillator coherent states: they are, respectively,
the minimum-uncertainty, the annihilation-operator, and the
displacement-operator method.

The states obtained by extension of these three
methods are in general different \cite{klauderskagilmore}.

The displacement-operator coherent states
are those preserving most of the properties of
the harmonic-oscillator coherent states: they are still overcomplete,
still enjoy resolution of unity, and moreover it was recently
discovered \cite{demartinoter94} that they follow classical
motion without dispersion by a dynamical feedback mechanism.

As to squeezed states, an extension of the minimum-uncertainty
and annihilation-operator methods to arbitrary non harmonic
systems was carried out by Nieto and
collaborators \cite{nieto79}, \cite{nieto93}.
They also introduced
an extension of the minimum-uncertainty method to obtain
generalized coherent states \cite{nieto78}.

However, an extension of the displacement-operator method
to obtain generalized squeezed states runs into difficulties
\cite{katriel} and is still missing.

We now show that the generalized ``dynamical" coherent states
defined in the previous section
via stochastic mechanics do in fact
define a particular class of
displacement-operator generalized squeezed states.

This is achieved by a natural extension of
the strategy that allowed to connect
the generalized coherent states introduced
in Ref.\cite{demartinobis94} via stochastic mechanics
with the displacement-operator generalized coherent states
\cite{demartinoter94}.

We proceed as follows. We first recall Glauber displacement
operator written in the coordinate representation:
\begin{equation}
\hat{D}_{\alpha} \; = \; \exp \left( iS_{0}(t)\right)
\exp \left( \frac{i}{\hbar}\langle \hat{p} \rangle \hat{q}
\right) \exp \left( -\frac{i}{\hbar}\langle \hat{q} \rangle
\hat{p} \right) \; .
\end{equation}

This operator, when applied to any
wave function $\Psi(x,t)$ displaces its space argument $x$
into $x - \langle \hat{q} \rangle $ and adds to its phase
the term $S_{0}(t) + \langle \hat{p} \rangle x$.

Next, we introduce a dynamical rescaling operator
$\hat{S}_{\Delta q}$, namely a squeeze operator defined as
\begin{equation}
\hat{S}_{\Delta q} \; = \;
\exp \left[ i \left( \frac{f(t)}{\hbar} \{ \hat{q}, \hat{p} \} +
\frac{g(t)}{(\Delta q_{0})^2}\hat{q}^{2} \right) \right] \, ,
\end{equation}

\noindent where
$\Delta q_{0}$ denotes the (time-independent!) ground state
dispersion. Given $\Delta q(t)$ solution of Eq. (22), the
two functions $f(t)$ and $g(t)$ read
\begin{equation}
f(t) \, = \, -\frac{1}{2}\ln \left( \frac{\Delta q}{\Delta q_{0}}
\right) \, , \; \; \; \; \; \; \; \; \; \;  g(t) \, = \, \frac{m}{\hbar}
(1-2f(t))^{-1}\frac{d}{dt}\ln \Delta q \, .
\end{equation}

\noindent We see from these relations that the
function $f(t)$ plays the role of a dynamical squeezing parameter.

Next, it is easily shown
that any ground state wave function can be
cast in the general form
\begin{equation}
\Psi_{0}(x) \; = \;
\frac{1}{\sqrt{\Delta q_{0}}} F \left( \frac{x}{\Delta
q_{0}} \right) \, ,
\end{equation}

\noindent with $F$ a suitably chosen function.

We now let $\hat{S}_{\Delta q}$ act on $\Psi_{0}$ to define
the dynamically rescaled wave function
\begin{equation}
\Psi_{resc}(x,t) \; = \; \hat{S}_{\Delta q} \cdot
\Psi_{0}(x) \, .
\end{equation}

Exploiting Trotter product formula and the algebra
of commutators, and observing that $\{ \hat{q}, \hat{p} \}
= i\hbar (1 + 2xd/dx)$ and that
$[\{ \hat{q} , \hat{p} \}
, \hat{q}^{2}] = -4i\hbar \hat{q}^{2}$,
one obtains
\begin{equation}
\Psi_{resc}(x,t) \; = \; \exp \left[ f(t) \left( 1 +
2x\frac{d}{dx} \right) \right] \cdot \chi(x,t) \, ,
\end{equation}

\noindent where $\chi(x,t)$ is given by
\begin{equation}
\chi(x,t) \; = \;
\exp \left[ i\frac{g(t)}{({\Delta q}_{0})^{2}}(1-2f(t))x^{2}
\right] \Psi_{0}(x) \, .
\end{equation}

We now exploit the extension to the real axis of the
following relation always true for analytic functions
(used in Ref. \cite{celeghini} in the context of
$q$-oscillators and coherent states):
\begin{equation}
Q^{x\frac{d}{dx}} \left[ W(x) \right] \; = \; W(Qx) \, ,
\end{equation}

\noindent with $Q$ any real $c$-number and $W$ any
function analytic on the real
axis. Letting $Q = \exp [2f(t)]$ and $W=\chi$
one finally is left with
\begin{equation}
\Psi_{resc}(x,t) \; = \; \exp \left[ f(t) +
i\frac{g(t)}{({\Delta q}_{0})^{2}}(1-2f(t)) e^{4f(t)}x^{2}
\right] \Psi_{0}\left( e^{2f(t)}x \right) \, .
\end{equation}

We then define the squeezed state $\Psi_{sq}$ by
applying $\hat{D}_{\alpha}$
on $\Psi_{resc}$:
\begin{equation}
\Psi_{sq}(x,t) \; = \; \hat{D}_{\alpha} \left( \hat{S}_{\Delta q}
\cdot \Psi_{0}(x) \right) \; = \; \hat{D}_{\alpha} \cdot
\Psi_{resc}(x,t) \, .
\end{equation}

By recalling Eqs. (28) it is then straightforward
to show that $\Psi_{sq}$ coincides
with the wave function Eq. (24) for the ``dynamical" coherent
states with classical motion and controlled dynamics
of the dispersion.
We thus proved that they can also be obtained
by a suitable extension of the
displacement-operator method.

{\it Concluding remarks}.
We have introduced a class of generalized coherent states
with time-dependent dispersion in the framework
of Nelson stochastic quantization.

The wave packets follow a classical evolution due to a
dynamical feedback, and the evolution
equation controlling the spreading of the wave packet
is naturally coupled with the classical evolution equation for
the wave packet centre.

We then showed that these states can be obtained
through a particular extension of
the displacement-operator method by defining a
dynamical rescaling operator.

The consequent dilatations or contractions of the wave
packet width are then shown to be controlled
by a single adimensional squeezing
parameter $f(t)$.

In this letter we outlined the general features of the
method: applications to specific potentials of
physical interest, as well as a more thorough discussion
of all the technical details will be given elsewhere
\cite{preparation}.

It might be worth noting that we
have chosen for simplicity a ground state to generate
squeezed states by the action of operators (26)-(27);
it is however immediately seen
that their application on any stationary state yields again
generalized coherent states of the form (24).

Finally, we remark that
Eq. (23) represents the ``stochastic squeezing" condition satisfied
by our states. Namely, it expresses the complementary
time-dependence of the spreading $\Delta q$ and
of the osmotic velocity
uncertainty $\Delta u$.

It is easily seen that in the canonic picture
Equations (10), (13), and (23) imply
$\Delta \hat q ^2 \Delta \hat p^2 = K + L^{2}(t)$, with $L(t)=m
\Delta \hat q d( \Delta \hat q )/dt$.

The reciprocal variation in time
of $\Delta \hat q$ and $\Delta \hat p$ is then ruled by
$\Delta \hat q$ itself, determined as the solution
of Eq. (22) with the initial condition $\Delta \hat q_{0}$.
In this way squeezing is introduced as a self-consistent
prescription on the dynamical evolution of the wave
packet spreading.

\end{document}